%% file: main.tex
\documentclass[sigconf]{acmart}

\AtBeginDocument{%
  }

\copyrightyear{2025}
\acmYear{2025}
\setcopyright{rightsretained}
\acmConference[CHIWORK '25]{CHIWORK '25: Proceedings of the 4th Annual Symposium on Human-Computer Interaction for Work}{June 23--25, 2025}{Amsterdam, Netherlands}
\acmBooktitle{CHIWORK '25: Proceedings of the 4th Annual Symposium on Human-Computer Interaction for Work (CHIWORK '25), June 23--25, 2025, Amsterdam, Netherlands}
\acmPrice{}
\acmDOI{10.1145/3729176.3729191}
\acmISBN{979-8-4007-1384-2/25/06}

\begin{document}

\title{AI, Jobs, and the Automation Trap: Where Is HCI?}


\author{Marios Constantinides}
\email{marios.constantinides@cyens.org.cy}
\orcid{0000-0003-1454-0641}
\affiliation{%
  \institution{CYENS Centre of Excellence}
  \city{Nicosia}
  \country{Cyprus}
  \postcode{1016}
}

\author{Daniele Quercia}
\email{quercia@cantab.net}
\orcid{0000-0001-9461-5804}

\affiliation{%
  \institution{Nokia Bell Labs}
  \city{Cambridge}
  \country{UK}
  \postcode{CB3 0FA}
}

\affiliation{%
  \institution{\vspace{1ex}Politecnico di Torino}  
  \city{Turin}
  \country{Italy}
}



\renewcommand{\shortauthors}{Constantinides and Quercia}

\begin{abstract}
As artificial intelligence (AI) continues to reshape the workforce, its current trajectory raises pressing questions about its ultimate purpose. Why does job automation dominate the agenda, even at the expense of human agency and equity? This paper critiques the automation-centric paradigm, arguing that current reward structures, which largely focus on cost reduction, drive the overwhelming emphasis on task replacement in AI patents. Meanwhile, Human-Centered AI (HCAI), which envisions AI as a collaborator augmenting human capabilities and aligning with societal values, remains a fugitive from the mainstream narrative. Despite its promise, HCAI has gone ``missing'', with little evidence of its principles translating into patents or real-world impact. To increase impact, actionable interventions are needed to disrupt existing incentive structures within the HCI community. We call for a shift in priorities to support translational research, foster cross-disciplinary collaboration, and promote metrics that reward tangible and real-world impact.
\end{abstract}



\maketitle

\input{sections/01_Introduction}
\input{sections/02_Patents}
\input{sections/03_HCAI}
\input{sections/04_Recommendations}
\input{sections/05_Conclusion}

\balance

\bibliographystyle{ACM-Reference-Format}
\bibliography{main}

\end{document}

%% file: sections/01_Introduction.tex
\section{Introduction}
\label{sec:introduction}

As artificial intelligence (AI) continues to permeate every facet of work, a fundamental question arises: What is its purpose in the workplace? Many AI initiatives today aim to automate tasks of occupations, often with the implicit goal of maximizing efficiency and reducing the need for human involvement~\cite{wef_jobs_llms_2025}. This trajectory has sparked concerns about widespread job displacement, reduced workforce agency, and exacerbation of existing societal inequalities.

Concerns about automation replacing human labor are not new. The Industrial Revolution offers a compelling historical parallel to today's AI-driven transformations. Just as 19\textsuperscript{th}-century workers resisted mechanization to preserve their livelihoods, today's workforce faces growing concerns about AI ``taking over'' human jobs~\cite{egan2024ai}. Reflecting on that period reveals a key lesson: \emph{when driven solely by efficiency, technological advancement can exacerbate inequality and displace vulnerable workers}. This historical lens sharpens our focus on the present challenge: \emph{will we allow AI to widen socioeconomic divides, or will we intentionally design systems that augment human work and promote inclusion?}

HCAI aims at doing the latter. It emphasizes the design of systems that align with human cognitive strengths, workflows, and societal values, empowering workers rather than marginalizing them~\cite{AIIndex2021}. 
Discussions in special interest groups (SIGs) at recent HCI conferences such as ACM CHI and CSCW~\cite{tahaei2023human, constantinides2024implications, deng2024collaboratively, zheng2024towards} have reinforced the critical need to align AI with human cognitive abilities. This alignment is evident in efforts such as adaptive learning platforms that personalize education for diverse learners~\cite{dutta2024enhancing}, and AI systems that augment radiologists to achieve greater diagnostic accuracy in collaboration with machine learning models~\cite{agarwal2023combining}.

However, despite decades of HCI work emphasizing participatory and human-centered design, the dominant trajectory of AI design and deployment remains centered on automation (i.e., replacing human tasks) rather than augmentation (i.e., supporting human capabilities). We critically examine this misalignment through  three research questions:
\vspace{-0.35cm}
\begin{itemize}
  \item[\textbf{RQ1}] \textbf{Does automation dominate AI’s current design ethos, despite long-standing alternatives?} We present empirical trends in AI patenting for workplace technologies to map automation-centric innovation trends (Section~\ref{sec:patents}).

  \item[\textbf{RQ2}] \textbf{Are alternatives fostering human augmentation possible?} We analyzed case examples of alternatives across HCAI's three application domains: education, healthcare, and workplace (Section~\ref{sec:hcai}). These examples are drawn from the literature based on conceptual relevance and application diversity rather than through a systematic review. 

  \item[\textbf{RQ3}] \textbf{How can HCI insights be better translated into real-world AI deployments?} We synthesized six actionable recommendations to enhance the translational impact of HCAI research (Section~\ref{sec:recommendations}).  These recommendations emerged from a reflective synthesis of our findings, and were informed by prior work in HCI~\cite{cao2023breaking}, HCAI~\cite{shneiderman2020human}, and Responsible AI~\cite{constantinides2024rai}.
\end{itemize}

%% file: sections/02_Patents.tex
\section{Does Automation Dominate AI’s current Design Ethos?}
\label{sec:patents}

\subsection{Automation Dominates}

\begin{figure*}
    \centering
    \includegraphics[width=0.99\linewidth]{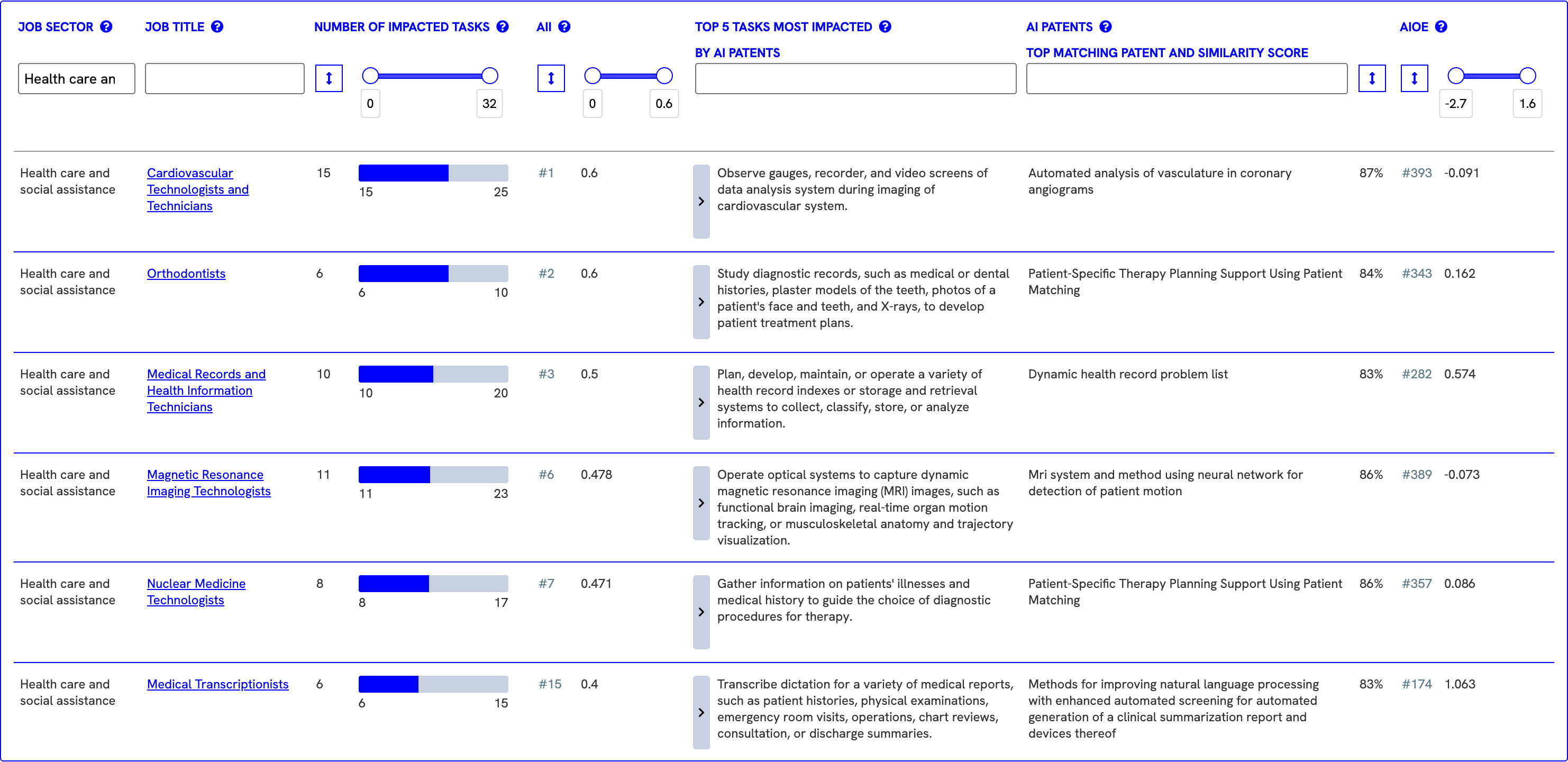}
    \caption{Jobs impacted by AI in the healthcare sector as visualized in the AI Impact dashboard (\url{https://social-dynamics.net/aii/}) based on the methodology described by Septiandri et al.~\cite{septiandri2024potential}. The most impacted jobs are primarily automation-driven, focusing on tasks such as automated vascular analysis and MRI system operations. In contrast, augmentation-driven tasks (e.g., patient-specific therapy planning and dynamic health record management) are less prominent.}
    \label{fig:aii_viz}
\end{figure*}

In two recent papers, we analyzed a dataset of 24,758 AI-related patents filed over the last decade~\cite{septiandri2024potential, kim2025potential}. This analysis was conducted  classifying patents according to whether they emphasized automation of job tasks or human involvement. Our goal was not to measure HCI's citation presence per se, but to evaluate the translational trajectory of AI design: do patents encode principles of augmentation or automation?

Our results showed that automation dominates over human augmentation in AI patents because the vast majority of patented AI innovations are aligned with task characteristics that favor replacement rather than collaboration~\cite{kim2025potential}. Specifically, AI patents reinforcing existing technologies and workflows are disproportionately associated with tasks that are predictable and performed individually, aligning with traditional models of automation. Even more disruptive AI patents primarily target mental and unpredictable tasks, yet still favor individual over collaborative work. In contrast, collaborative tasks remain largely untouched by either type of AI, suggesting that augmenting human interaction and cooperation remains a marginal concern in patentable AI development. 

To gain a more nuanced and in-depth understanding of how AI patents conceptualize labor, we turned to healthcare as a representative sector. The selection of this domain was based on its rich intersection with both automation and augmentative technologies (e.g., AI is extensively used to automate diagnostic and administrative tasks such as analyzing medical imaging~\cite{Forbes2024, van2024adapted}, predicting patient outcomes~\cite{american2023ama}, and managing documentation~\cite{liu2022hipal}), and because it is a high-stakes domain where the impact of automation versus augmentation has immediate human consequences. 

A strong emphasis on automation tasks over augmentation is present in healthcare too (Figure~\ref{fig:aii_viz}). Tasks such as automated analysis of medical imaging (e.g., coronary angiograms and MRI scans), dominate the landscape, as seen with patents focused on technologies such as `dynamic magnetic resonance imaging' or `automated vascular analysis'. In contrast, fewer patents target augmentation technologies that assist medical professionals directly (e.g., systems for patient-specific therapy planning or dynamic health record problem lists), which involve supporting rather than replacing human expertise. These findings are in line with recent publications. A report by the American Medical Association found, in fact, that 65\% of physicians recognized the utility of AI in reducing administrative burdens, and over half highlighted its potential in diagnostic support~\cite{american2023ama}. Similarly, a Forbes report noted that AI systems are predominantly applied to automate processes such as disease detection and treatment planning, emphasizing efficiency and scalability over collaboration and augmentation~\cite{Forbes2024}. However, while augmentation-focused applications such as clinical decision support and personalized medicine exist, they represent a smaller portion of AI's integration into healthcare~\cite{american2023ama}.

\subsection{Why Does Automation Dominate?}

One may now wonder why the dominant focus of AI innovation remains on designing systems that operate autonomously. We offer four complementary explanations.

\vspace{1em}
\noindent
\textbf{Because augmenting humans is not easy.} Human-centered systems must carefully balance the benefits of AI with concerns around job satisfaction, meaning, and professional identity~\cite{ghosh2024impact}. For instance, Toner-Rodgers~\cite{toner2024artificial} found that pairing researchers with AI systems led to substantial productivity gains: scientists discovered 44\% more materials, and patent filings increased by 39\%. However, 82\% of them also reported feeling less fulfilled in their work. This highlights a paradox: \emph{while AI can boost efficiency and innovation, it may inadvertently diminish the intrinsic rewards of creative and intellectual engagement}~\cite{bainbridge1983ironies}.

\vspace{1em}
\noindent
\textbf{Because long-standing alternatives stem from a field with limited translational impact.} HCI research, despite offering robust frameworks for participatory and human-centered design, has historically struggled to influence industry practices. Only 13.4\% of publications from SIGCHI-sponsored venues (e.g., CHI, CSCW, UIST, UbiComp) have been cited in AI patents, compared to 25\% from fields such as computer vision~\cite{cao2023breaking}. This points to a translational gap: a disconnect between academic innovation and its real-world application. While HCI contributions such as direct manipulation~\cite{shneiderman1983direct}, heuristic evaluation~\cite{nielsen1990heuristic}, social translucency~\cite{erickson2000social}, and value-sensitive design~\cite{friedman1996value,sadek24} have shaped theory and practice, they often remain absent from patent-driven innovation ecosystems.

\vspace{1em}
\noindent
\textbf{Because automation aligns with dominant worldviews.} Historically, men have filed most patents~\cite{jensen2018gender} and continue to receive the majority of AI funding~\cite{turing2024rebalancing}, embedding particular economic values such as efficiency and control into emerging technologies. As Wajcman argues~\cite{wajcman2007women}, dominant perspectives in technological development often prioritize performance metrics over relational or affective dimensions of work. As a result, the very notion of ``success'' in AI-human collaboration is shaped by gendered assumptions about productivity rather than a more inclusive view of  human flourishing.

\vspace{1em}
\noindent
\textbf{Because automation aligns with short-term economic incentives.} The reward structures governing AI innovation typically favor short-term gains: reducing labor costs, scaling operations, and maximizing shareholder value. Yet this economic logic often comes at the expense of long-term societal outcomes. Autor~\cite{autor2024new, Autor2024} has shown that automation disproportionately targets routine, middle-skill jobs, contributing to job polarization and middle-class erosion. While automation boosts productivity, it narrows pathways to economic mobility. Without intentional investment in technologies that augment human labor, the current trajectory risks exacerbating inequality~\cite{autor2024applying}.

%% file: sections/03_HCAI.tex
\section{Any Alternatives Fostering  Augmentation?}
\label{sec:hcai}

To address the automation-centric trajectory identified in the previous section, we turn to examples of Human-Centered AI (HCAI) systems that already support human expertise, decision-making, and collaboration. Rather than removing human agency, these systems exemplify the shift toward augmentation by helping individuals stay in control, build new skills, and make informed decisions~\cite{shneiderman2020human, jorke2024supporting}. Building on prior work~\cite{constantinides2024good}, we analyzed three key domains\footnote{The examples in this section are not intended as a systematic review but were selected through a semi-structured search process. We reviewed recent HCI and AI literature with a focus on augmentative design, practical deployment, and cross-sectoral representation. The final selection aimed to balance conceptual diversity, domain relevance, and empirical richness, while illustrating how HCAI principles are applied in practice.} that are central to the structure of modern labor markets: \emph{education}, \emph{workplace}, and \emph{healthcare}. Education prepares future workers; workplace technologies support those currently in the labor force; and healthcare represents a high-stakes sector undergoing rapid AI-driven change. These domains also reflect many of the 138 mobile and wearable AI use cases recently evaluated against the EU AI Act for their risks and benefits~\cite{constantinides2024good}. Across all three, we observed how HCAI principles are being applied to build systems that promote trust, enhance human capability, and serve as viable alternatives to automation-focused deployments.

\subsection{Augmenting Learning and Development in Education}
Recent studies presented at ACM CHI have showcased innovative educational tools that leverage AI for personalized learning. Although education is not a work setting per se, it plays a foundational role in preparing individuals for a labor market increasingly shaped by AI. Adaptive learning platforms, for example, dynamically adjust content to meet individual student needs that significantly improve engagement and learning outcomes. For example, Cheng et al.~\cite{cheng2024scientific} demonstrated the potential of augmented reality (AR) and AI-driven systems to create immersive and culturally relevant learning experiences. Narrative-driven AR applications, coupled with LLMs, could enhance both cognitive and emotional development in children by fostering deep engagement with educational content. Similarly, GPT Coach leverages LLMs to provide tailored tutoring experiences~\cite{jorke2024supporting}. By incorporating motivational interviewing techniques and leveraging data from wearables, GPT Coach exemplifies how HCAI principles can create personalized and supportive learning environments. Further, GPTeach~\cite{markel2023gpteach} introduced a novel approach to teacher training by utilizing LLM-powered simulated students to create scalable and low-risk environments for novice educators. Through interactive teaching scenarios, GPTeach enables teachers to refine their instructional strategies, offering immediate feedback and fostering reflective practice. Such tools exemplify the transformative potential of HCAI in education. By personalizing interactions and fostering agency, they empower students and educators alike to achieve deeper learning outcomes and cultivate skills tailored to their individual needs and aspirations. The integration of AR, narrative-based pedagogy, and LLM-driven systems provides a robust pathway for augmenting human capabilities, shifting the focus from replacing educators to augmenting them.

\subsection{Augmenting Human Capabilities in the Workplace}
AI-driven tools in the workplace are transforming how employees engage with complex tasks, enabling real-time feedback and fostering collaboration. For example, GitHub Copilot and similar code recommendation systems significantly enhance programmer productivity by offering real-time coding suggestions and reducing the cognitive load associated with repetitive tasks~\cite{mozannar2024reading}. Although AI coding assistants have been shown to expedite task completion for experienced programmers~\cite{moassessing}, yet their effects on novice programmers remain debated. Some evidence suggests these tools may hinder learning by bypassing foundational understanding~\cite{gardella2024performance}. Designing such systems with pedagogical scaffolding and explainability could better support skill development and augmentation in programming tasks. Intelligent systems designed for online meetings further augment collaboration by addressing communication challenges in virtual settings. For example, MeetCues leverages real-time visual and interactive feedback to enhance communication in virtual meetings~\cite{aseniero2020meetcues, zhou2022predicting}, and KAIROS uses multimodal monitoring (e.g., body language, speech patterns) through wearable devices to capture non-verbal communication cues that otherwise might go unnoticed during virtual meetings.  In addition to software engineering and productivity tools, AI is augmenting decision-making in sectors such as finance and law. In finance, AI dashboards assist financial analysts by flagging anomalies and synthesizing data streams to support rather than replacing expert judgment~\cite{gomber2018fintech}. In the legal domain, AI-powered tools help summarize documents and draft briefs that enable faster iteration while maintaining professional oversight~\cite{cao2022ai}. These systems exemplify how AI can offload repetitive tasks while preserving human responsibility and ethical accountability~\cite{alfrink2023contestable}.

\subsection{Augmenting Decision-Making in Healthcare}
AI systems such as diagnostic assistants analyze patient data to provide clinicians with actionable insights, helping to improve decision-making in high-stakes environments. For example, it has been shown that adapted LLMs can outperform medical experts in clinical text summarization tasks, including summarizing radiology reports, patient questions, progress notes, and doctor–patient dialogue~\cite{van2024adapted}. As LLM-generated summaries were rated as equivalent or superior to those created by medical experts in terms of completeness, correctness, and conciseness, by integrating such models into clinical workflows could alleviate the documentation burden and allow clinicians to allocate more time to patient care. However, as Autor~\cite{Autor2024} highlights, the effectiveness of such tools depends on the ability of healthcare professionals to understand and utilize them appropriately. For example, a recent study by Nikhil Agarwal et al. demonstrated that while AI diagnostic tools performed at least as accurately as nearly two-thirds of radiologists in an experimental setting, they did not improve the quality of diagnoses. The key issue was that radiologists often misinterpreted or misused the AI's predictions. Confident AI predictions were frequently overridden by clinicians, and when AI expressed uncertainty, doctors often relied on these predictions even when their own assessments were better~\cite{agarwal2023combining}. Research evidence also suggests that AI tools can help reduce burnout among healthcare professionals by automating routine tasks and alleviating cognitive load~\cite{liu2022hipal}. Centered on augmentation rather than replacement, this example illustrates how HCAI principles can guide the development of AI systems that enhance both patient care and healthcare worker well-being.

%% file: sections/04_Recommendations.tex
\section{How Can HCI Translate into Real-world AI Deployments?}
\label{sec:recommendations}

To translate the principles of HCAI into real-world impact, we must align ethical imperatives with institutional incentives. In today's political and business climate, where efficiency, scalability, and profitability often dominate, the success of augmentative AI systems depends not only on moral alignment, but also on strategic advantage. Based on our review of patent trends, case studies, and the broader HCI literature, we propose six recommendations to increase the translational impact (Table~\ref{tab:recommendations}). These recommendations are not prescriptive blueprints, but synthesized themes that emerged from cross-domain analysis and our own practitioner experience in HCI research and design. We present them here as entry points for discussion and iteration.
\smallskip

\begin{table*}[t!]
\centering
\caption{Summary of recommendations for increasing HCAI's translational impact.}
\label{tab:recommendations}
\scalebox{0.95}{
\begin{tabular}{p{4cm}|p{5cm}|p{5cm}}
\hline
\textbf{Recommendation} & \textbf{Key Actions} & \textbf{Expected Impacts} \\ \hline
Responsible AI by Design & 
- Incorporate ethical guidelines (e.g., fairness, transparency) into AI workflows. \newline
- Use tools like interactive dashboards to monitor adherence to principles. & 
- Ensures AI systems align with regulatory frameworks (e.g., EU AI Act). \newline
- Reduces risks of bias and enhances public trust. \\ \hline
Fostering Academia-Industry Collaboration & 
- Develop partnerships with industries such as healthcare and manufacturing. \newline
- Establish research labs embedded in companies to align with real-world needs. & 
- Facilitates faster adoption of HCAI innovations. \newline
- Informs research agendas based on practical workflows. \\ \hline
Developing Scalable Prototypes for Workforce Applications & 
- Create job-specific AI prototypes addressing key industry needs (e.g., healthcare diagnostics). \newline
- Make prototypes available as open-source tools to encourage broader use. & 
- Provides clear examples of augmentative AI in action. \newline
- Encourages collaboration and faster scaling across industries. \\ \hline
Targeting Job Preservation and Skill Development & 
- Design AI systems to complement human labor in automation-prone sectors. \newline
- Develop adaptive learning platforms for reskilling and upskilling workers. & 
- Mitigates job displacement by preparing workers for AI-augmented roles. \newline
- Addresses labor shortages in caregiving, education, and logistics. \\ \hline
Embedding Feedback Loops to Address Worker Concerns & 
- Use participatory design to include workers in system development. \newline
- Implement contestability mechanisms to enable workers to challenge AI decisions. & 
- Ensures systems align with worker needs and reduce physical or cognitive strain. \newline
- Increases trust and satisfaction with AI tools. \\ \hline
Shaping Policy for Augmentative AI & 
- Collaborate with government agencies to prioritize augmentation in AI policies. \newline
- Advocate for regulations requiring participatory design and human oversight. & 
- Aligns AI deployment with societal values like fairness and inclusivity. \newline
- Encourages policies that reduce job polarization. \\ \hline
\end{tabular}
}
\end{table*}

\subsection{Taking a ``Responsible AI by Design'' Approach} 
``Responsible AI by Design'' involves building AI systems responsibly from the very beginning, starting in the design phase, to prevent issues later in development. Developers and industry leaders should prioritize safety from the outset, taking inspiration from how self-driving car companies integrate safety measures at every stage of their design process. This approach ensures that AI aligns with human needs, integrates seamlessly into daily life, minimizes risks, and remains productive and helpful. By adopting a structured approach~\cite{constantinides2024rai}, developers can establish clear Responsible AI (RAI) guidelines. These guidelines, grounded in regulations such as the EU AI Act, help ensure that safety and fairness are embedded throughout the development process~\cite{yfantidou2023beyond}. A practical way to implement this is by incorporating RAI guidelines into AI workflows using interactive tools and visual dashboards. For example, digital prompts or cards that highlight ethical considerations at each stage of development can promote adherence to core principles such as fairness, transparency, and safety~\cite{elsayed2023responsible, eccolaCards2021}. While ethical guidelines such as fairness and transparency are essential, they must also be framed in terms of operational and reputational benefits. Research shows that responsible design mitigates costly downstream failures, reduces compliance burdens, and strengthens public trust~\cite{constantinides2024rai, elsayed2023responsible, eccolaCards2021, liao2020questioning}; factors that increasingly affect brand perception and market share. As the regulatory landscapes evolve (e.g., EU AI Act), adopting these principles early can serve as a strategic safeguard against reputational or legal harm.

\subsection{Fostering Academia-Industry Collaboration for Workforce Innovation}
To ensure HCAI research addresses real-world job needs, researchers should prioritize partnerships with industries where AI is already transforming work (e.g., manufacturing, healthcare, and service sectors)~\cite{septiandri2024potential, autor2024new}. Joint initiatives, such as workforce retraining programs or augmented workplace studies, can inform research agendas and facilitate faster adoption of innovations. Establishing research labs embedded within companies could help HCAI researchers gain deeper insights into job workflows, enabling them to design AI systems that genuinely augment human labor. While establishing in-house research labs may appear costly, the long-term return on investment can be substantial. Embedded collaboration helps companies avoid deployment failures by aligning AI systems with actual work practices, reducing workforce resistance, and enhancing employee retention. Similar to how usability testing reduces product returns and support costs, HCI-informed augmentation strategies can reduce turnover and burnout while improving system adoption. As Burke et al.~\cite{burke2010social} show, systems designed with social and well-being outcomes in mind can yield broader engagement benefits. Furthermore, evidence from organizational research suggests that collaborative and participatory design processes improve trust, which is vital for AI adoption in high-stakes domains~\cite{vaccaro2024combinations}.

Cross-disciplinary collaborations between HCI researchers, social scientists, economists, and policymakers are also essential~\cite{wallach2025position}. For example, partnerships with labor economists can provide evidence of the economic benefits of augmentation, while collaborations with policymakers can ensure these insights influence regulatory frameworks. Future work in this area could focus on making these guidelines easier to use in real-world AI applications, developing tools that help developers create AI systems that enhance rather than replace human jobs, and studying how these guidelines impact employment opportunities and workforce dynamics across different industries. Additionally, as academic research often emphasizes novelty over practical impact, it is important to develop new reward structures that incentivize researchers to perform translational work~\cite{cao2023breaking, septiandri2024impact}.

\subsection{Developing Scalable Prototypes for Workforce Applications}
HCAI researchers should focus on developing scalable and job-specific AI prototypes that offer clear benefits by augmenting workers' capabilities. Ideally, these prototypes should be made available as open-source tools to encourage broader adoption and collaboration. For example, ExploreGen~\cite{herdel2024exploregen} demonstrates how generative AI can support researchers in quickly creating prototypes while simultaneously identifying potential risks and benefits, ensuring alignment with human-centered goals and responsible design principles. Such prototypes can include AI tools that assist mid-skill workers, like diagnostic support systems for medical technicians, enabling them to make better decisions.  Finally, AI systems designed to reduce cognitive load can streamline complex tasks for professionals in high-stakes environments, such as air traffic controllers or radiologists, enhancing both accuracy and efficiency~\cite{herdel2024exploregen, constantinides2024good}.

\subsection{Empowering Workers Through Job Preservation and Skill Development}
AI systems that disproportionately automate routine tasks have contributed to the erosion of middle-skill jobs, leading to job polarization and economic inequality~\cite{acemoglu2022tasks, septiandri2024potential, goos2009job}. In roles vulnerable to automation, AI systems should complement human labor rather than replacing it. However, we acknowledge that in certain high-risk environments (e.g.,  manufacturing or chemical processing) automation can increase worker safety by removing people from hazardous conditions. For example, replacing human labor on factory floors with AI-driven machines can prevent accidents involving heavy machinery or toxic exposure. Yet even in these contexts, hybrid systems that maintain human oversight or offer augmentation through robotics-assisted safety mechanisms can strike a more balanced approach between efficiency and worker agency. Another way would be to integrate skill development initiatives into the design and deployment of these systems, ensuring workers are equipped to transition into higher-value tasks within their existing roles. For example, AI-driven training simulators and adaptive learning tools can help workers acquire skills tailored to emerging demands in industries such as manufacturing and healthcare~\cite{jagannathan2019dominant}. These systems can bridge the gap between traditional education and augmented workflows, preparing workers to thrive in AI-augmented roles. Additionally, by collaborating with educational institutions, HCAI researchers can promote AI literacy and integrate vocational training programs that address the specific needs of AI-impacted sectors. Moreover, addressing labor shortages in critical areas such as caregiving and education requires designing AI systems that enhance efficiency without compromising the essential human qualities of empathy and creativity. For example, augmentation tools that assist caregivers with administrative tasks can enable them to focus on delivering personalized care. While previous literature has suggested reallocating low-skilled workers to tasks requiring creative and social intelligence~\cite{frey2017future}, we argue that (re)skilling and upskilling efforts should be approached cautiously, ensuring that new AI systems align with workers' capabilities and societal needs. For example, while a truck driver may not transition to an electronics technician, they might excel as a logistics coordinator or sales professional. Similarly, a logistics coordinator transitioning from truck driving can benefit from AI-driven scheduling systems that enhance decision-making without requiring extensive retraining. At the same time, opportunities for (re)skilling and upskilling can also foster joy, self-efficacy, and personal development; factors that are critical for long-term worker engagement and well-being. Future work should explore the diverse aspirations and capacities of workers across industries to design augmentation-focused interventions that resonate with their needs.

\subsection{Embedding Feedback Loops to Address Worker Concerns}
To align HCAI research with job-centric challenges, embedding workers and managers into the feedback loops of AI development processes is essential. This approach ensures that AI systems not only meet technical requirements but also address the practical needs and concerns of those who interact with them daily. A key component of such integration is the concept of AI contestability, which emphasizes designing systems that allow decisions and processes to be challenged or adapted by those directly or indirectly affected~\cite{alfrink2023contestable}. For example, involving warehouse workers in the participatory design of robotics-assisted picking systems ensures that these technologies align with ergonomic needs, reducing physical strain and improving productivity. Similarly, engaging healthcare workers in designing AI-assisted diagnostics tools fosters systems that fit seamlessly into clinical workflows, minimizing disruption and improving user satisfaction~\cite{verma2023rethinking}. Such participatory methods create a collaborative feedback loop where stakeholders can contest and refine system decisions, leading to tools that are both effective and ethically aligned with human needs. Another way of allowing workers to understand AI's impact is through visualization tools that highlight disparities caused by automation (e.g., job displacement rates across industries or demographic groups).

\subsection{Shaping Policy for Augmentative AI}
HCAI researchers can significantly increase their impact by shaping policy frameworks that guide the deployment of AI in the workforce. Recent insights from special interest groups organized at ACM CHI and CSCW have highlighted how regulatory policies such as the EU AI Act can support augmentative AI systems by mandating transparency, user empowerment, and the ability for individuals to contest AI decisions~\cite{tahaei2023human, constantinides2024implications, deng2024collaboratively}. These policies align with Human-Centered AI principles, emphasizing the need for AI systems to augment human labor rather than replace it.  However, influencing policy requires HCI to consider policy engagement not as peripheral but as central to its intellectual and practical concerns. As recent work suggests, HCI can expand its role by positioning system-people-policy interactions at the core of its research, practice, and education~\cite{yang2024future}. This approach allows HCI to blend system, human, and policy expertise seamlessly, creating a cohesive framework for addressing the societal implications of AI. Such a shift enables the HCAI community to move beyond isolated policy engagements and work collectively to influence outcomes that prioritize human augmentation. Collaborating with government agencies, labor unions, and industry stakeholders, HCAI researchers can advocate for policies that encourage participatory design and ensure AI systems enhance worker well-being. For example, promoting mandates that require human oversight in AI-driven decision-making processes can foster trust and usability while mitigating risks associated with automation~\cite{alfrink2023contestable, yurrita2023disentangling}. Furthermore, HCI researchers can provide empirical evidence on the benefits of augmentative AI, drawing on case studies where AI has successfully enhanced human capabilities in healthcare, education, and other critical sectors. 

%% file: sections/05_Conclusion.tex
\section{Conclusion}
\label{sec:conclusion}

Our analysis demonstrates a systemic disconnect between HCI's augmentation-oriented ethos and AI’s current automation-first trajectory. While HCAI has emerged as a compelling framework, the path from design principle to industrial deployment remains unclear. Part of the challenge lies in incentive structures: patents, funding models, and commercial benchmarks often reward cost-reduction and task elimination rather than human empowerment or resilience.

Moreover, the line between automation and augmentation is blurry. In some domains, automation may in fact enable augmentation by, for example, reducing cognitive load or freeing time for complex tasks. However, this requires intentional design. Without proactive interventions, automation risks degrading work quality or removing workers from critical feedback loops, thereby increasing system fragility.

The recommendations outlined in this paper, ranging from embedding feedback loops to advocating for augmentative AI policies, highlight actionable pathways for increasing HCAI's translational impact. We proposed strategies that emphasize collaboration with diverse stakeholders, including workers, educational institutions, and policymakers, to ensure that AI aligns with human values and societal needs. By doing so, HCAI will be positioned not only as a moral imperative, but as a competitive advantage.

\section*{AI USAGE DISCLOSURE STATEMENT}
In the writing and revision stages, we used a generative AI tool (i.e., OpenAI's ChatGPT-4o) to support activities such as refining the clarity of our arguments and enhancing overall readability. The tool was used as a writing aid and did not generate original scientific content or research findings.